\documentclass[%
 reprint,
superscriptaddress,
 amsmath,amssymb,
 aps,
]{revtex4-1}

\usepackage{amssymb, amsmath} 
\usepackage{graphicx}
\usepackage{xspace}
\usepackage{color}
\usepackage{url}
\usepackage{dcolumn}
\usepackage{bm}

\usepackage{setspace}

\begin{document}

\preprint{APS/123-QED}

\title{First muon acceleration using a radio frequency accelerator}

\newcommand{\KEK}{\affiliation{High Energy Accelerator Research Organization (KEK), Tsukuba, Ibaraki 305-0801, Japan}}
\newcommand{\JAEA}{\affiliation{Japan Atomic Energy Agency (JAEA), Tokai, Naka, Ibaraki 319-1195, Japan}}
\newcommand{\JPARC}{\affiliation{J-PARC Center, Tokai, Naka, Ibaraki 319-1195, Japan}}
\newcommand{\Tokyo}{\affiliation{University of Tokyo, Hongo, Tokyo 171-8501, Japan}}
\newcommand{\Ibaraki}{\affiliation{Ibaraki University, Mito, Ibaraki 310-8512, Japan}}
\newcommand{\Nagoya}{\affiliation{Nagoya University, Nagoya, Aichi 464-8602, Japan}}
\newcommand{\Riken}{\affiliation{Riken, Wako, Saitama 351-0198, Japan}}
\newcommand{\SNU}{\affiliation{Seoul National University, Seoul 08826, Republic of Korea}}
\newcommand{\INPA}{\affiliation{Institute for Nuclear and Particle Astrophysics, Seoul National University, Seoul, 08826, Republic of Korea}}
\newcommand{\KU}{\affiliation{Korea University, Seoul 02841, Republic of Korea}}
\newcommand{\BINP}{\affiliation{Budker Institute of Nuclear Physics, SB RAS, Novosibirsk 630090, Russia}}
\newcommand{\NSU}{\affiliation{Novosibirsk State University, Novosibirsk 630090, Russia}}
\newcommand{\Pulkovo}{\affiliation{Pulkovo Observatory, St. Petersburg, 196140, Russia}}

\author{S. Bae}
\author{H. Choi}
\author{S. Choi}\SNU\INPA

\author{Y. Fukao}
\author{K. Futatsukawa}\KEK

\author{K. Hasegawa}\JAEA

\author{T. Iijima}\Nagoya

\author{H. Iinuma}\Ibaraki

\author{K. Ishida}\Riken

\author{N. Kawamura}\KEK

\author{B. Kim}\SNU\INPA

\author{R. Kitamura}\Tokyo

\author{H. S. Ko}\SNU\INPA

\author{Y. Kondo}\email{yasuhiro.kondo@j-parc.jp}\JAEA 

\author{S. Li}\Tokyo

\author{T. Mibe}
\author{Y. Miyake}\KEK

\author{T. Morishita}\JAEA

\author{Y. Nakazawa}\Ibaraki

\author{M. Otani}\email{masashio@post.kek.jp}\KEK

\author{G. P. Razuvaev}\BINP\NSU\Pulkovo

\author{N. Saito}\JPARC

\author{K. Shimomura}\KEK

\author{Y. Sue}\Nagoya

\author{E. Won}\KU

\author{T. Yamazaki}\KEK


\begin{abstract}
Muons have been accelerated by using a radio frequency accelerator
for the first time.
Negative muonium atoms (Mu$^-$), which are bound states of positive
muons ($\mu^+$) and two electrons, are generated from $\mu^+$'s
through the electron capture process in an aluminum degrader.
The generated Mu$^-$'s are initially electrostatically accelerated and
injected into a radio frequency quadrupole linac (RFQ).
In the RFQ, the Mu$^-$'s are accelerated to 89~keV.
The accelerated Mu$^-$'s are identified by momentum measurement
and time of flight.
This compact muon linac opens the door to various muon accelerator
applications including particle physics measurements and the construction 
of a transmission muon microscope.
\end{abstract}

\maketitle

\section{Introduction}
\label{sec:intro}

Since its invention, the radio frequency (RF) accelerator has 
accelerated a wide variety of particles from electrons to
rare isotopes, and greatly contributed to the progress of various
branches of science.
Recently, the demand for muon acceleration has arisen not only
in the field of elementary particle physics, but also in 
material and life sciences.
For example, in muon collider and neutrino
factory studies~\cite{palmer:neutrino_factory:icfa_bd_newsletter2011},
it is proposed that the large transverse emittance of the muon beam
can be reduced using ionization
cooling~\cite{neuffer:principles_muon_cooling:part_accel1983}. 
A muon beam passes through a material, and subsequently 
the lost energy in the material is restored using RF acceleration. 
After all the cooling processes, muons are accelerated from a few
MeV with RF accelerators.
In material and life sciences, one promising
application of muon acceleration is in the construction of
a transmission muon microscope.
If the muons can be cooled to the thermal temperature
(ultraslow muon, USM) and subsequently re-accelerated,
transmission muon microscopes will be
realized~\cite{miyake:transmission_muon_microscope}.
The remarkable progress made with modern proton drivers enables 
the USM generator to be used as a particle source of accelerators.
Because the mass of the muon is 200 times larger than that of the
electron, the transmission depth of a 10~MeV muon reaches 
approximately 10~$\mu$m.
This enables three-dimensional imaging of living cells, 
which is impossible with the use of transmission electron microscopes.
Another application of USM acceleration is precise measurement 
of the muon anomalous magnetic moment $a_\mu = (g-2)_\mu / 2$ 
and electric dipole moment (EDM).
Muon acceleration is essential to realize these applications; 
however, it has not been demonstrated except for simple electrostatic 
acceleration.
In this letter, the first demonstration of muon RF acceleration is
presented.
It was conducted during the development of the muon RF linear accelerator (linac)
for a muon g-2/EDM experiment.

\section{A g-2/EDM experiment at J-PARC}
\label{sec:g-2}

The $(g-2)_\mu$ anomaly is one of the most promising methods to
explore physics beyond the Standard Model of elementary particle
physics.
Currently, the most precise measurement of $a_\mu$ has been performed
by the E821 experiment of Brookhaven National
Laboratory~\cite{bennett:e821:prd2006}.
The precision is 0.54~ppm, and the measured value is approximately
three standard deviations from the Standard Model
prediction~\cite{hagiwara:g-2_reevaluated:jps2011, davier:reavaluation_g-2:epj2011, benayoun:update_g-2:epj2013}.
To improve this precision, a new muon g-2/EDM experiment is proposed 
at Japan Proton Accelerator Research Complex (J-PARC).
This experiment, J-PARC E34, aims to measure $a_\mu$ with a precision of
0.1~ppm and the EDM with a precision of
$1 \times 10^{-21}\ e \cdot$cm~\cite{e34_cdr}.
Unlike E821 and its predecessors and
successor~\cite{e989_tdr}, E34 will use a low-emittance muon beam.
The required transverse momentum spread $\Delta p_{t}/p$ is
less than $10^{-5}$, and the assumed transverse emittance is
1.5~$\pi$~mm~mrad.
To satisfy this requirement, the 25 meV USMs generated by laser
dissociation of the thermal muoniums (Mu, or $\mu^+$e$^-$) from
a silica aerogel
target~\cite{beer:enhancement_of_muonium_emission:ptep2014}
will be accelerated to 212~MeV using a 
muon linac~\cite{otani:develop_mulinac:linac2016}.
The muon linac will be constructed at the
H line~\cite{kawamura:h_line:usm2013} of the J-PARC muon science
facility (MUSE)~\cite{miyake:muse:jpc2010}.
It will consist of a radio frequency quadrupole linac
(RFQ)~\cite{kondo:simulation_mu_rfq:ipac2015},
an inter-digital H-mode drift tube
linac~\cite{otani:ih_dtl_design:prab2016},
disk-and-washer coupled cell linac~\cite{otani:develop_mulinac:ipac2016},
and disk loaded traveling wave
structures~\cite{kondo:beamdynamics_muon_highbeta:ipac2017}.

The muon linac is an unproven technology, and thus muon acceleration should be
demonstrated as soon as possible prior to the construction of
the actual linac.
In addition, because the intensity of the muon linac is much lower
than that of an ordinary linac, a commissioning method should be
established through a muon acceleration experiment.
The laser-dissociation USM source is now being developed with high
priority;
however, an earlier and simpler slow muon source is necessary to conduct
the muon acceleration experiment.
The scheme of muon cooling using a simple metal degrader and
re-acceleration with an RFQ, originally proposed at Los Alamos National
Laboratory~\cite{miyadera:design_of_muon_accelerator:pac2007},
is suitable for this purpose.
We basically follow this method, but the emittance of the simply
degraded muon is too large compared to the RFQ acceptance.
Therefore, we use epithermal negative muoniums 
(Mu$^-$, or $\mu^+$e$^-$e$^-$) generated from degraded $\mu^+$'s 
through the electron capture
process~\cite{kuang:formation_negative_muonium:pra1989}.
The emittance of this Mu$^-$ beam is still larger than the RFQ
acceptance, but much smaller than that of the degraded $\mu^+$'s.
Moreover, the energy spectrum of simply degraded $\mu^+$'s is very
broad, reaching up to the full RFQ accelerated energy.
Some of those penetrate the RFQ without acceleration.
We call these $\mu^+$'s penetrating $\mu^+$'s.
This makes it difficult to distinguish them from the accelerated
$\mu^+$'s.
In contrast, the Mu$^-$'s have a sharp peak near zero energy,
and the accelerated Mu$^-$'s are easily separated from the
penetrating $\mu^+$'s because they have opposite charge.
Using this slow Mu$^-$ source and a prototype RFQ of the J-PARC
linac~\cite{kondo:fabrication_50mArfq:linac2006},
we conducted the muon acceleration experiment in a multi-purpose
experimental area (D2 area) of MUSE.

\section{Experimental Apparatus}
\label{sec:apparatus}

\begin{figure*}[!hbt]
\centering
\includegraphics[width = 0.95\textwidth]{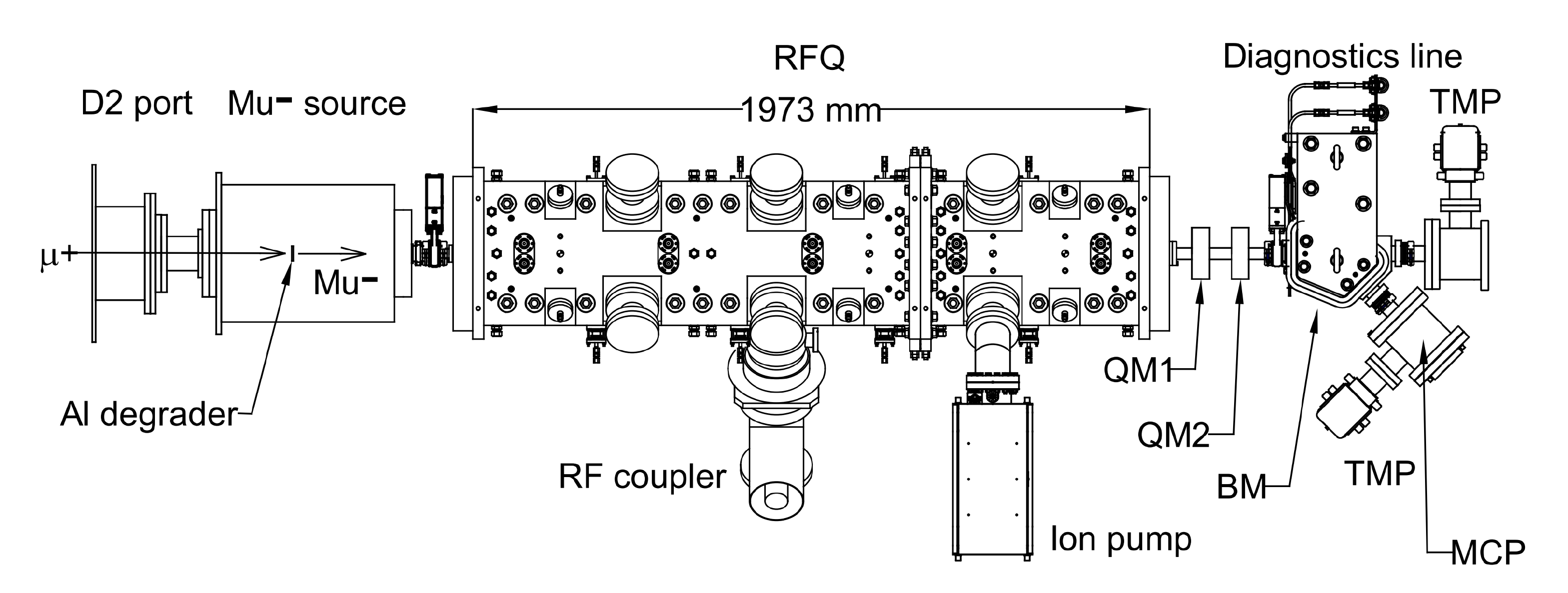}
\caption{Schematic drawing of the experimental setup.}
\label{fig:setup}
\end{figure*}

Figure~\ref{fig:setup} shows a schematic drawing of the experimental
setup.
The MUSE facility provides a pulsed surface muon ($\mu^{+}$) beam
produced by $\pi^+$ decay near the surface of the production target.
The beam pulse width is 47~ns in rms, and
the repetition rate is 25~Hz~\cite{strasser:decay_muon_channel:jpc2010}.
For this experiment, the beam power of the J-PARC Rapid Cycling
Synchrotron (RCS) was 300~kW.
The energy of the $\mu^+$ beam was chosen to be 2.9~MeV to maximize
the Mu$^-$ emission yield on the basis of a separate
experiment~\cite{kitamura:first_trial_muon_acc:ipac2017}, which 
was conducted prior to this acceleration experiment,
and individually measured $\mu^+$ deceleration and Mu$^-$ production
itself.
With this beamline setting, the $\mu^+$ intensity was estimated to be
$3 \times 10^6$~/s.
The $\mu^+$'s were incident on an aluminum degrader with dimensions
$43 \times 40$~mm$^2$ and a thickness of 200~$\mu$m.
The beam profile of the $\mu^+$ at the Al degrader was estimated
from a beamline simulation based on the G4beamline simulation
package~\cite{g4beamline}.
The estimated profile was verified by compasiron with 
the measured profile at the $\mu^+$ focal point of the D2 area.
The position of the Al degrader is different from the $\mu^+$ focal point
due to a geometrical constraint, and therefore the horizontal and vertical
rms beam sizes at the Al degrader were estimated from the simulation to
be 32~mm and 28~mm, respectively;
thus, only 27\% of the primary $\mu^+$ hit the Al degrader.
The rms energy spread is estimated to be 16\%.
The $\mu^+$'s were decelerated through the Al degrader, and some 
$\mu^+$'s captured two electrons to become Mu$^-$'s at the 
downstream surface of the Al degrader.
The conversion efficiency from $\mu^+$ to $\mathrm{Mu^{-}}$ 
is estimated to be $8\times10^{-7}$ from  
the data of the separate
experiment~\cite{kitamura:first_trial_muon_acc:ipac2017}. 
By using an electrostatic lens system, a Soa
lens~\cite{Canter:Positron_Studies_of_Solids},
the generated Mu$^-$'s were accelerated to 5.6~keV and focused
on the entrance of the RFQ.

As mentioned above, the prototype RFQ of the J-PARC linac was
used for this experiment.
The length of this RFQ corresponds to two thirds of the length of
the complete 3~MeV RFQ, and this prototype RFQ was designed to 
accelerate negative hydrogen ions (H$^-$'s) up to 0.8~MeV.
This RFQ employs a conventional beam dynamics design;
that is, the inter-vane voltage $V$ and the average bore 
radius are maintained constant except for the injection section.
The cell parameters were designed with
KEKRFQ~\cite{ueno:new_beam_dynamics:linac1990}, and the number of
the cells is 297.
In order to use this RFQ for muon acceleration, the voltage $V$ should
be normalized to the muon mass, and the input velocity $\beta$ should
be the same as that of the H$^-$, as shown in Table~\ref{tbl:conversion}.

\begin{table}[htp]
   \centering
   \caption{RFQ parameter conversion from H$^-$ to $\mu$.}
   \begin{tabular}{lll}
       \hline
           Particle \hspace{3cm}   & H$^-$ \hspace{5mm} & $\mu$ \hspace{5mm}\\
           Mass (MeV/c$^2$)        & 939.3 & 105.7             \\
           Injection $\beta$       & \multicolumn{2}{c}{0.010} \\
           Injection energy (keV)  & 50    & 5.6               \\
           Extraction $\beta$      & \multicolumn{2}{c}{0.041} \\
           Extraction energy (keV) & 810   & 89                \\
           Inter-vane voltage (kV)  & 80.7  & 9.08              \\
           Nominal power (kW)      & 180   & 2.3               \\
       \hline
   \end{tabular}
   \label{tbl:conversion}
\end{table}

\begin{figure}[hbt]
\centering
\includegraphics[width = 0.45\textwidth, bb = 0 0 691 518]
                {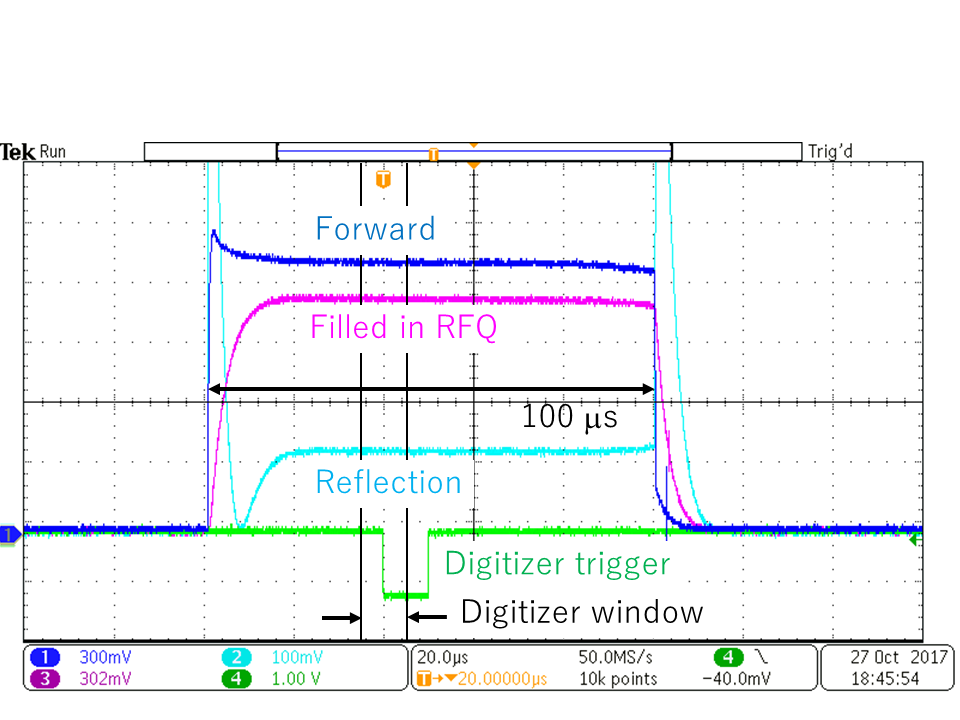}
\caption{RF pulses input (Forward) into, reflected from, and filled
         in the RFQ. The digitizer window for the beam detector is also
         shown.}
\label{fig:rf_waveform}
\end{figure}

This RFQ is longitudinally separated into two modules.
Each module consists of upper and lower major vanes and left and 
right minor vanes made of oxygen-free copper.
The vane tips and inside surfaces of the cavity were machined
with formed cutters.
The machining accuracy of the vane tips was within 15~$\mu$m.
In the center port of the lower-right quadrant, a loop-type RF coupler 
was inserted.
The RF power was set to 2.3~kW for the muon acceleration operation,
and transmitted via 50-$\Omega$ coaxial cables.
The forward and reflected RF powers were measured through a directional
coupler attached to the RF coupler, and the RF power in
the RFQ cavity was measured using a loop pickup monitor inserted
into the cavity.
Figure~\ref{fig:rf_waveform} shows the typical RF waveform of the RFQ.
In this figure, the RF pulses input into (Forward) and reflected from
(Reflection) the RFQ, the filled power in the RFQ, and the trigger for
the waveform digitizer of the beam detector are indicated.
The RF pulse width was 100~$\mu$s and the repetition rate was 25~Hz.
 
The extracted beam properties were measured using a beam diagnostics
line.
The beam was transferred using two quadrupole magnets (QM1 and QM2)
with field gradients of 2.6~T/m and 1.8~T/m, respectively. 
The charge and momentum of the particle can be selected with a
bending magnet (BM).
In our experiment, because the intensity of the Mu$^-$ is very low,
it is not feasible to find the correct field setting with
the ordinary BM-current scanning method.
Instead, we used H$^-$'s generated by exposing the Al degrader
surface to ultraviolet light.
If the acceleration voltage of the Soa lens is set to 10~kV,
the momentum of the H$^-$'s is the same as that of the 89~keV Mu$^-$'s.
The field setting of the BM was verified using 10~keV H$^-$
prior to the muon acceleration experiment.

A microchannel plate (MCP, Hamamatsu photonics F9892-21~\cite{hamaphoto})
was located at the downstream end of the $45^\circ$ line.
The fiducial area of the MCP detector corresponded to a circle of
42~mm diameter, and the aperture ratio of the micro channel was 60\%.
The MCP waveform, in an interval of 10~$\mu$s around each 25-Hz beam pulse, 
was digitized with a 250~MS/s waveform digitizer.
The $\pm 5~\mu s$ from the leading edge of the digitizer trigger in
Fig.~\ref{fig:rf_waveform} corresponds to the digitizing window.
A pulse higher than the pedestal level of the waveform was regarded
as a signal pulse.
The leading edge of the signal pulse and the maximum height within
the signal window of 40~ns were defined as the signal timing and
pulse height, respectively. 

\section{Result}
\label{sec:results}

\begin{figure}[hbt]
\centering
\includegraphics[width=0.5\textwidth]{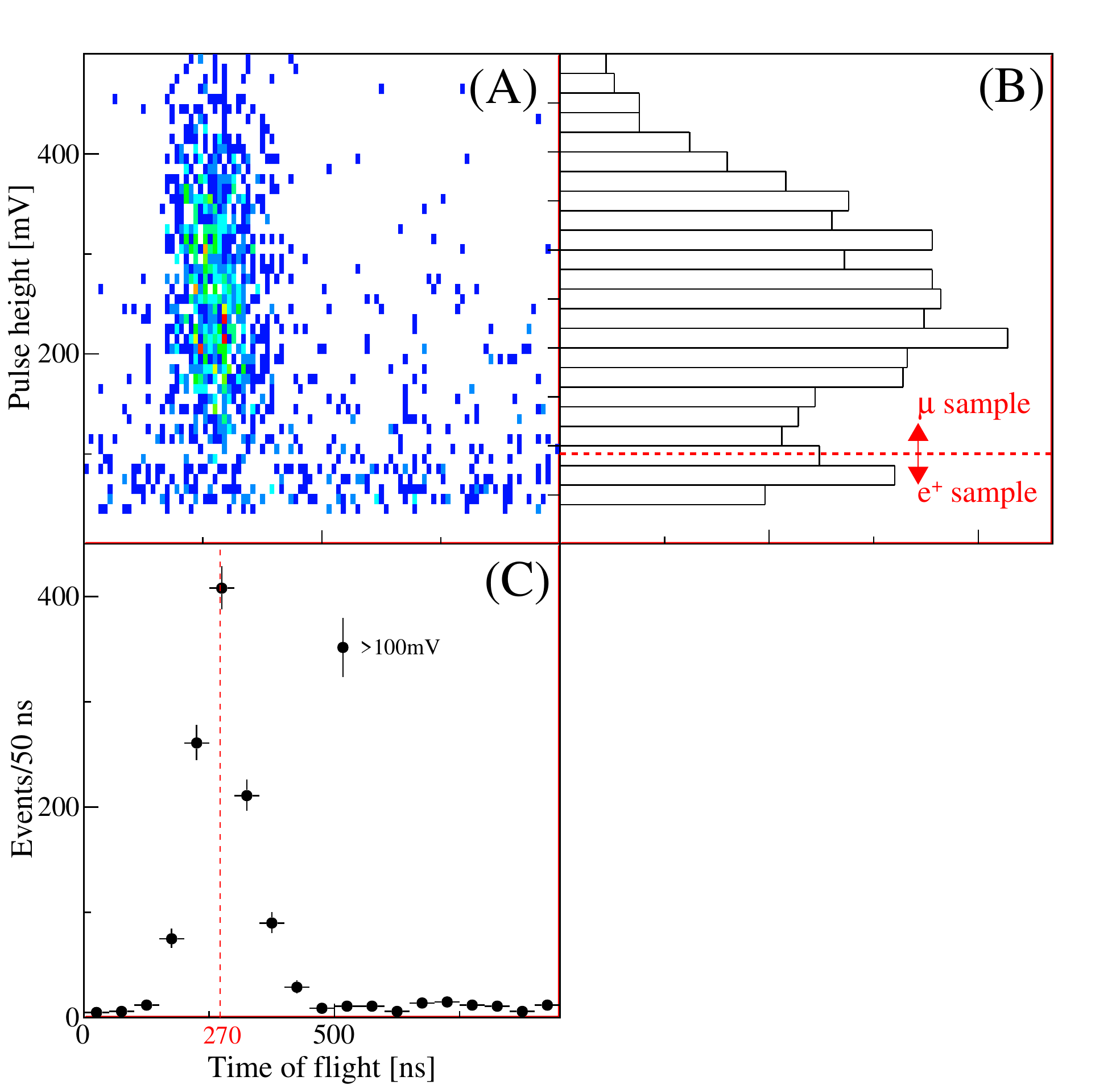}
\caption{Distribution of the MCP pulse height and the
  TOF of the penetrating $\mu^+$.
  (A) Scatter plot of the pulse height vs TOF.
  (B) Pulse height of the MCP signal. The events above 100~mV were
  regarded as $\mu^+$.
  (C) TOF spectrum after the pulse-height cut was applied.
  The peak corresponds to the $\mu^+$'s injected into the RFQ with
  an energy of 89~keV.}
\label{fig:tof_vs_plheight}
\end{figure}

At the beginning of the muon acceleration experiment,
the beam diagnostics system was verified using penetrating $\mu^+$'s.
The energy spectrum of the penetrating $\mu^+$'s is continuous up to
more than 90 keV, and $89 \pm 18$~keV $\mu^+$'s can be selected with
the BM. 
Figure~\ref{fig:tof_vs_plheight}(A) is a scatter plot of pulse height 
vs. time of flight (TOF) for the observed $\mu^+$ with the MCP. 
The muon arrival time at the Al degrader was measured with a set of
scintillating counters located at the side of the Al degrader. 
Figure~\ref{fig:tof_vs_plheight}(B) is a projection to the
pulse-height axis.
The main background of the muon measurement is decay positrons from
the $\mu^+$, but they penetrate the MCP, and thus are easily
eliminated by applying a pulse-height cut.
The threshold was determined from Figs.~\ref{fig:tof_vs_plheight}(A)
and (B) as 100~mV.
Figure~\ref{fig:tof_vs_plheight}(C) shows the TOF distribution
after the pulse-height cut was applied. 
Because the distance between the Al degrader and the MCP is
3.4 m, the TOF of the 89~keV $\mu^+$ is 270~ns.
The observed TOF peak is consistent with this calculation. 
The rms timing width of 50~ns mainly comes from that of the primary
muon beam at the Al degrader.

\begin{figure}[hbt]
\centering
\includegraphics[width=0.5\textwidth]{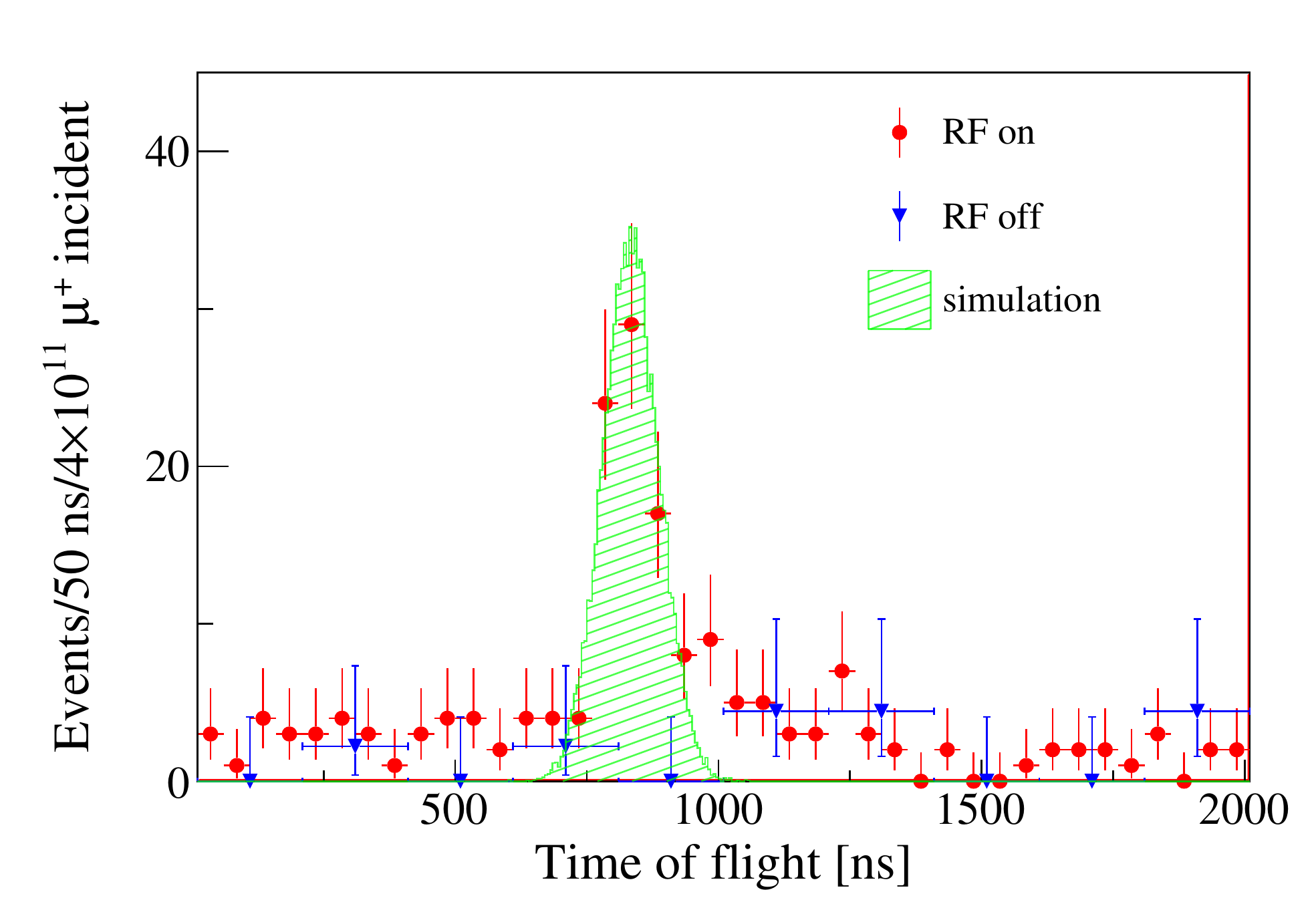}
\caption{TOF spectra of the negative-charge configuration with RF on and off.
  The clear peak of the RF on spectrum at 830~ns corresponds to
  the accelerated Mu$^-$'s. The error bars are statistical.
  A simulated TOF spectrum of the accelerated Mu$^-$'s is also plotted.}
\label{fig:tof}
\end{figure}

Finally, the polarities of the magnets were flipped to
the negative-charge configuration.
Figure~\ref{fig:tof} shows the TOF spectrum with and without
the RF operation after the pulse-height cut was applied.
With the RF operation, a clear peak was observed at $830\pm11$~ns. 
The error is the statistical error of the peak position of the Gaussian
fitting. 
The number of cells of this 324-MHz RFQ is 297, and thus it takes
$\frac{297}{2\times324\times10^6} = 458$~ns to fully accelerate the
particles through the 324~MHz RFQ.
Therefore, the arrival time of the accelerated Mu$^-$ is later than
that of the penetrated $\mu^+$.

The TOF spectrum was confirmed with a series of simulations.
The simulation of the Soa lens was conducted using GEANT4~\cite{geant4}. 
The three-dimensional electric field was calculated with
OPERA3D~\cite{opera3d} and implemented in the simulation. 
The transit time through the Soa lens was estimated with this simulation
to be 307~ns and the acceptance was estimated to be 4\%. 
PARMTEQM~\cite{parmteq} was employed for the RFQ simulation, and 
the transmission was estimated to be 5\%. 
Almost all losses occurred at the RFQ entrance,
because of much larger emittances than the acceptance of the RFQ. 
TRACE3D~\cite{trace3d} and PARMILA~\cite{parmila} were utilized for the
diagnostics line simulation. 
The transport efficiency to the MCP was evaluated to be 87\%. 
The length of the diagnostics line is 0.91~m, and thus the transit time
of the 89~keV Mu$^-$ is 72~ns.
The total flight time of the accelerated Mu$^-$  
from the Al degrader to the MCP was calculated to be
$t_{\mathrm{tran.}} = 307 + 458 + 72 = 837$~ns, which is consistent with
the measurement.
The hatched histogram in Fig.~\ref{fig:tof} represents the simulated 
TOF spectrum of the accelerated Mu$^-$. 
The number of simulation events was normalized to 
$4 \times 10^{11}$ incident $\mu^+$'s.
The muon survival rate was calculated to be
$\exp(t_{\mathrm{tran.}} / \tau_{\mu} ) = 81$\%. 
The 46~ns rms width of the TOF spectrum is consistent with that
from the timing distribution of the primary $\mu^+$ at the Al degrader.

From these experimental results, it is concluded that the observed
TOF peak is due to the Mu$^-$'s accelerated by the RFQ to 89~keV.
The event rate in the 780 to 980~ns TOF range was estimated to be
$(5\pm1) \times 10^{-4}$/s by subtracting the decay-positron events
estimated from the timing region outside the signal range. 

\section{Conclusion}
\label{sec:conclusion}

In summary, muons have been accelerated by RF acceleration
for the first time.
Slow negative muonium atoms (Mu$^-$) were generated through the electron
capture process of the degraded $\mu^+$'s in the D2 area of
J-PARC MUSE, and accelerated with the RFQ up to 89~keV.

The intensity of the accelerated Mu$^-$ in this experiment is limited by 
the very low conversion efficiency of 2.9~MeV $\mu^+$ to Mu$^-$.
With the construction of the new H line and assuming the design value
of 1-MW beam power from the RCS, the intensity is expected to be
$2 \times 10^{-2}$/s.
Structures to further accelerate the beam from the RFQ are now
being developed, and can be demonstrated using this beam.
There is also the possibility of significantly improving
the conversion efficiency by cesiation as with H$^-$ ion
sources~\cite{dudnikov:cold_muonium_production:ipac2017}.
Finally, the laser-dissociation ultraslow muon source
is expected to be installed in this beamline to obtain
a muon rate of $10^6$/s, the design intensity for the g-2/EDM 
experiment.
The result presented in this letter is the first step toward making
the low-emittance muon beam available as a powerful tool for 
application in material and life sciences and fundamental physics research. 

\begin{acknowledgments}

The authors would like to thank the J-PARC muon section staffs for their
support in the conduct of the experiment at J-PARC MUSE.
We also thank the J-PARC and KEKB linac groups for their support
in assembling the accelerator components.
We express our appreciation to the many manufacturing companies
involved in this project, especially to Toshiba Co., who fabricated
the RFQ.
This work is supported by JSPS KAKENHI Grant Numbers
JP25800164
, JP15H03666
, JP16H03987
, JP15H05742
, and JP16J07784. 
This work is also supported by
the Korean National Research Foundation grants
NRF-2015H1A2A1030275,
NRF-2015K2A2A4000092,
and NRF-2017R1A2B3007018;
the Russian Foundation for Basic Research
grant RFBR 17-52-50064; 
and the Russian Science Foundation grant
RNF 17-12-01036.
This experiment at the Materials and Life Science Experimental Facility
of the J-PARC was performed under user programs (Proposal No.
2017A0263).

\end{acknowledgments}



\hyphenation{Post-Script Sprin-ger}
%

\end{document}